# O eterno retorno hoje

Juliano Neves*

**Resumo**: Neste trabalho, realizamos uma aproximação entre as cosmologias científicas não singulares (sem a singularidade inicial, o *big bang*), especialmente os modelos cíclicos, e o pensamento do eterno retorno de Nietzsche. Além disso, sugerimos respostas para o porquê da busca por "provas" científicas do eterno retorno feita pelo filósofo na década de 1880.
**Palavras-chave**: ciência - cosmologia - eterno retorno – forças

## 1 – O pensamento mais elevado e abissal

Em *Ecce homo*, Nietzsche escreve que "(...) o *pensamento do eterno retorno*, a mais elevada forma de afirmação que se pode em absoluto alcançar, é de agosto de 1881: foi lançado em uma página com o subscrito: 'seis mil pés acima do homem e do tempo'" (EH/EH, *Assim falou Zaratustra* 1, KSA 6.335. Trad. PCS[1]). Pensamento elevado, o eterno retorno também é o mais profundo, o mais abissal, pois conduz à visão da eterna repetição sem sentido ou fim de tudo. Sua importância não é pequena na filosofia nietzschiana. Na obra publicada, aparece pela primeira vez no *A gaia ciência* como desafio e hipótese:

> O mais pesado dos pesos. — E se um dia ou uma noite um demônio se esgueirasse em tua mais solitária solidão e te dissesse: "Esta vida,

---

\* Doutor em Física pelo Instituto de Física da Universidade de São Paulo, São Paulo, Brasil. E-mail: nevesjcs@if.usp.br.
1 Indicamos os tradutores da obra de Nietzsche em português pelas iniciais dos seus nomes e sobrenomes nas passagens citadas: PCS (Paulo César de Souza), RRTF (Rubens Rodrigues Torres Filho) e FRK (Flávio R. Kothe). As traduções feitas das referências em inglês são de minha responsabilidade.

assim como tu vives agora e como a viveste, terás de vivê-la ainda uma vez e ainda inúmeras vezes; e não haverá nela nada de novo, cada dor e cada prazer e cada pensamento e suspiro e tudo o que há de indivisivelmente pequeno e de grande em tua vida há de te retornar, e tudo na mesma ordem e sequência (...)". Não te lançarias ao chão e rangerias os dentes e amaldiçoarias o demônio que te falasse assim? Ou viveste alguma vez um instante descomunal, em que lhe responderias: "Tu és um deus, e nunca ouvi nada mais divino!" Se esse pensamento adquirisse poder sobre ti, assim como tu és, ele te transformaria e talvez te triturasse; a pergunta, diante de tudo e de cada coisa: "Quero isto ainda uma vez e ainda inúmeras vezes?" pesaria como o mais pesado dos pesos sobre o teu agir! (FW/GC 341, KSA 3.570. Trad. RRTF).

A ideia de um tempo circular, de um eterno retorno, não é nova[2]. Mesmo Nietzsche afirmou que a doutrina do eterno retorno "(...) poderia afinal ter sido ensinada também por Heráclito. Ao menos encontram-se traços dela no estoicismo, que herdou de Heráclito quase todas as suas ideias fundamentais" (EH/EH, *O nascimento da tragédia* 3, KSA 6.313. Trad. PCS). Mas em Nietzsche a doutrina serve como condição para a superação do "mais estranho e mais ameaçador de todos os hóspedes"[3]: o niilismo. O eterno retorno, então, coloca-se como nova referência suprema, novo peso[4], depois da morte de Deus. "É o novo 'peso', a concepção de uma eternidade que se realiza no tempo, capaz de mudar todo o âmbito dos valores", escreve Rubira[5]. Nessa interpretação, o eterno retorno é a condição necessária para a transvaloração de todos os valores. Assim, "o mais pesado dos pesos", título do aforismo do *A gaia ciência*, tem duplo sentido: como o maior dos pesos, o eterno retorno sufocará o ressentido, o fraco, aquele que não suporta a vida e, muito menos, sua repetição eterna; para o homem afirmador, aquele que aprova a vida em sua totalidade, o eterno retorno do mesmo será a condição para criar valores. Por outro lado, além do aspecto ético, podemos pensar o eterno retorno do mesmo em Nietzsche também como uma cosmologia[6]. Num fragmento póstumo, o filósofo aponta nessa direção:

> E sabeis sequer o que é para mim "o mundo"? (...) Este mundo: uma monstruosidade de força, sem início, sem fim, uma firme, brônzea grandeza de força, que não se torna maior, nem menor, (...) jogo de forças e ondas de forças ao mesmo tempo um e múltiplo, (...) afirmando ainda a si próprio, nessa igualdade de suas trilhas e anos, abençoando a si próprio como Aquilo que eternamente tem de retornar (...) (Nachlass/FP 1885, 38[12], KSA 11.610. Trad. RRTF).

---

[2] De acordo com Eliade, "no terceiro século a. C, Berossus popularizou a doutrina caldeia do 'Grande Ano', numa forma que se espalhou por todo mundo helênico (de onde passou depois para os romanos e os bizantinos). Segundo essa doutrina, o universo é eterno, mas sofre uma destruição periódica, sendo reconstituído a cada Grande Ano (…). É bastante provável que esta doutrina de periódicas conflagrações universais também fosse defendida por Heráclito (…)." Cf. ELIADE, M. *Mito do eterno retorno*. Trad. José A. Ceschin. São Paulo: Mercuryo, 1992, p.78.

[3] "O niilismo está diante da porta: de onde nos chega esse mais estranho e mais ameaçador de todos os hóspedes?" (Nachlass/FP 1885-86, 2[127], KSA 12.125. Trad. FRK).

[4] Cf. Rubira. L. *Nietzsche: do eterno retorno do mesmo à transvaloração de todos os valores.* São Paulo: Discurso Editorial/Editora Barcarolla, 2010. Rubira trata do significado da palavra peso, em Nietzsche, como sinônimo de valor, especialmente no capítulo 2.

A cosmologia do eterno retorno do mesmo nietzschiana é uma interpretação do mundo como "Aquilo que eternamente tem de retornar", sempre "na mesma ordem e sequência". É esse aspecto no pensamento de Nietzsche — o cosmológico — que tratamos neste trabalho, aproximando a tese ou doutrina nietzschiana às cosmologias científicas atuais.

---

5 Ibidem, p. 317.
6 Cf. MARTON, S. O eterno retorno do mesmo: tese cosmológica ou imperativo ético. In: *Extravagâncias: ensaios sobre a filosofia de Nietzsche*. 3. ed. São Paulo: Discurso Editorial/Editora Barcarolla, 2009.

## 2 – A ciência na cosmologia nietzschiana

Como escreve Marton: "Por muito tempo, e por razões diversas, estudiosos da filosofia de Nietzsche pretenderam ignorar que suas preocupações foram ditadas, por vezes, muito mais pelas questões da investigação científica de sua época que pelos problemas filosóficos ou filológicos (...)."[7] Pelo atual número crescente de trabalhos publicados[8] que relacionam o pensamento nietzschiano e as ciências, principalmente as naturais, vemos que essa relação não foi pequena e desprezível. Em *Nietzsche: das forças cósmicas aos valores humanos*, Marton enfatiza a importância dos trabalhos dos biólogos Rolph e Roux para a criação do conceito de vontade de potência: "Ora, em 1881, de Roux, Nietzsche reteve a noção de que, no próprio organismo, entre órgãos, tecidos e células, existe concorrência vital e, em 1884, de Rolph, a noção de que a concorrência, em vez de prejudicar a vida, aumenta sua quantidade"[9]. A comentadora acrescenta que "no conceito de vontade de potência, as duas noções serão subsumidas."[10] Em Boscovich, físico, matemático e astrônomo croata, que viveu no século XVIII, Nietzsche foi buscar o conceito de força, importante para sua cosmologia. Segundo o físico Max Jammer, Boscovich, um estudioso do fenômeno de colisão dos corpos, chegou à conclusão que "impenetrabilidade e extensão (…) são meramente expressões espaciais de forças, 'força' é consequentemente mais fundamental do que matéria (…)."[11] Nietzsche compreendeu Boscovich e escreveu em *Além*

---

7 Cf. MARTON, S. Da biologia à física: vontade de potência e eterno retorno do mesmo. Nietzsche e as ciências da natureza. In: *Nietzsche e as ciências*, org. Miguel Barrenechea *et al*. Rio de Janeiro: Editora 7 Letras, 2011, p. 114.
8 A coletânea *Nietzsche e as ciências*, organizada por Barrenechea *et al.*, levanta essa importante questão.
9 Cf. MARTON, S. *Nietzsche: das forças cósmicas aos valores humanos*. 3. ed. Belo Horizonte: Editora UFMG, 2010, p. 64.
10 Idem, Ibidem.
11 Cf. JAMMER, M. *Concepts of force*. New York: Dover Publications Inc., 1999, p. 178.

*do bem e do mal*: "(...) Boscovich nos ensinou a abjurar a crença na última parte da terra que permanece firme, a crença na 'substância', na 'matéria', nesse resíduo e partícula da terra, o átomo (...)" (JGB/BM 12, KSA 5.26. Trad. PCS).

Mas ao conceito de força da mecânica newtoniana, Nietzsche acrescentou um caráter intrínseco, um mundo interior: a vontade de potência. "O vitorioso conceito de 'força', com o qual nossos físicos recriaram Deus e o mundo, necessita ainda de uma complementação: é necessário aditar-lhe um mundo interior, ao qual eu chamo de 'vontade de potência'"[12] (Nachlass/FP 1885, 36[31], KSA 11.563. Trad. FRK). Num outro fragmento póstumo, o filósofo reforça essa ideia: "Uma força que não podemos conceber (como a assim chamada força de atração e de repulsão puramente mecânica) é uma palavra vazia (...). Todo acontecer derivado de propósitos é redutível ao *propósito de ampliar a potência*" (Nachlass/FP 1885-86, 2[88], KSA 12.105. Trad. FRK). Se em *Assim falou Zaratustra* a vontade de potência aparece somente ligada à vida[13] ou ao mundo orgânico, posteriormente, na obra nietzschiana, aparece relacionada ao mundo inorgânico também: *"Esse mundo é a vontade de potência — e nada além* disso!" (Nachlass/FP 1885, 38[12], KSA 11.611. Trad. RRTF). O mundo como forças em luta, em disputa, buscando mais potência — essa é a cosmologia nietzschiana. Forças que, na totalidade, não aumentam ou diminuem, forças que se efetivam numa eternidade imanente, um mundo que "eternamente tem de retornar":

> Se o mundo *pode* ser pensado como grandeza determinada de força e como número determinado de centros de força — e toda outra representação permanece indeterminada e consequentemente *inutilizável* —, disso se segue que ele tem de passar por um número calculável de combinações, no grande jogo de dados de sua existência. Em um tempo infinito, cada combinação possível estaria alguma vez alcançada; mais ainda: estaria alcançada infinitas vezes (Nachlass/FP 1888, 14[188], KSA 13.376. Trad. RRTF).

---

12  Nas traduções de PCS e FRK, substituímos a expressão "vontade de poder" por "vontade de potência". Além disso, nas traduções de FRK, preferimos força ao invés de energia. Isso se deve porque consideramos que com o uso do conceito de energia perde-se a noção de luta e disputa.

13  "Apenas onde há vida há também vontade: mas não vontade de vida, e sim — eis o que te ensino — vontade de potência!" (Za/ZA II, *Da superação de si* mesmo, KSA 4.149. Trad. PCS).

Por serem finitas ou determinadas, as forças repetem seus arranjos num tempo infinito. Tudo "na mesma ordem e sequência" retornaria infinitas vezes. A cosmologia nietzschiana funda-se numa eternidade imanente e no número finito de forças. O retorno de tudo, segundo o filósofo, segue necessariamente.

Algumas tentativas de "provar" ou mostrar o absurdo da tese nietzschiana foram realizadas. Em *Nietzsche: sua filosofia dos antagonismos e os antagonismos de sua filosofia*, Müller-Lauter discute[14] algumas dessas "provas" ou "contraprovas", como as de Becker e Simmel. Mas nesse trabalho, uma possível "prova" da tese cosmológica ficou fora da discussão: o teorema do eterno retorno do matemático Henri Poincaré. Já Paolo D'Iorio[15] e Stephen Brush[16] fazem uma aproximação entre a tese nietzschiana e esse poderoso teorema matemático.

Publicado em 1890, o teorema do eterno retorno afirma que um sistema mecânico finito com quantidade finita de energia deve eventualmente retornar arbitrariamente próximo a um estado inicial qualquer num número infinito de vezes. O teorema de Poincaré, segundo Brush, é uma tentativa de contestar a visão materialista e mecanicista de mundo, especialmente a teoria cinética dos gases.

---

14  As discussões sobre as "provas" e "contraprovas" encontram-se no capítulo 7.
15  Cf. D'IORIO, P. The eternal return: genesis and interpretation. In: *Agonist: a Nietzsche circle journal*, vol. 4, 2011. A menção ao teorema de Poincaré é feita no epílogo
16  Cf. BRUSH, S. *The kind of motion we call heat: a history of kinetic theory of gases in the 19th century*. Amsterdam: Elsevier Science & Publishers B. V., 1976, vol. II, p. 628. Nessa parte do livro, o autor de forma inusitada faz um paralelo entre as visões do retorno de Nietzsche e Poincaré e afirma que o último provou matematicamente as intuições do primeiro.

Partindo de leis mecânicas, o teorema conduz à conclusão da reversibilidade de processos mecânicos. Assim, inicialmente, o teorema do eterno retorno mostrou duas opções: ou abandono da segunda lei da termodinâmica, que aponta para a irreversibilidade de processos físicos, ou o abandono da descrição mecânica do mundo (o que Poincaré tinha em mente). Mas, de acordo com Brush, o físico Ludwig Boltzmann — que tinha uma visão estatística da segunda lei da termodinâmica — argumentou "que a propriedade de retornar não é um paradoxo, mas uma consequência perfeitamente razoável da concepção estatística da matéria. Podemos computar a probabilidade de uma flutuação estatística que retornaria o sistema ao seu estado inicial; mas essa probabilidade, enquanto finita, é tão pequena que o retorno somente ocorreria em eras incontáveis de tempo."[17] Na interpretação de Boltzmann, o teorema do eterno retorno e a segunda lei da termodinâmica não eram incompatíveis.

Na sequência do fragmento 14[188] de 1888 acima citado, Nietzsche, sem ter tido a possibilidade de conhecer o teorema de Poincaré (Nietzsche teve a paralisia progressiva em 1889), afirma equivocadamente que a concepção do eterno retorno "não é, sem mais, uma concepção mecanicista: pois se fosse, não condicionaria mais um infinito retorno de casos idênticos, e sim um estado final." Como já dissemos, segundo o teorema de Poincaré, uma descrição inteiramente mecânica do mundo pode conduzir à reversibilidade, ao retorno de tudo.

Na próxima seção, tratamos de "provas", ou melhor, indicações que corroboram a ideia de um eterno retorno do mesmo usando *apenas* as cosmologias científicas atuais não singulares, aquelas que eliminam o *big bang* (a grande explosão).

---

17  Ibidem, p. 96.

Neves, J.

*3 – As cosmologias hoje*

O modelo padrão da cosmologia científica descreve, atualmente, um universo em expansão há 13,7 bilhões de anos. São três os seus ingredientes: (1) as equações de Einstein, ou seja, a teoria geométrica da gravitação einsteiniana, a relatividade geral, publicada em 1916; (2) o princípio cosmológico, que afirma a igualdade de todos os observadores no universo (num dado instante, em diferentes posições, todos os observadores têm a mesma interpretação do cosmo); (3) acima de uma certa escala de distância, quando o universo apresenta-se homogêneo e isotrópico[18], o conteúdo de matéria do universo é descrito por um fluído perfeito. O primeiro ingrediente nos diz como matéria e energia comportam-se diante da geometria e vice-versa; o segundo traz consigo ecos de um ideal moderno (aqui, a igualdade não vale apenas na Terra, vale em qualquer parte do universo!); o terceiro, uma simplificação útil e funcional. Mesmo sendo muito bem sucedido, o modelo padrão enfrenta problemas como o pouco conhecimento sobre a natureza da matéria escura e da energia escura. A primeira é um tipo de matéria que altera, por exemplo, a velocidade de rotação das galáxias; a segunda, um tipo exótico de energia que torna a expansão do universo cada vez mais acelerada. Ambas têm sido estudadas em diferentes abordagens: na cosmologia padrão, no entanto, a matéria escura é fria, não emite e nem absorve luz, e a energia escura é representada nas equações de Einstein pela famosa constante cosmológica[19].

---

18   Um espaço-tempo isotrópico tem as mesmas propriedades em qualquer direção, não há uma direção privilegiada.
19   A constante cosmológica teria sido considerada pelo próprio Einstein o seu maior erro. Quando formulou sua teoria da gravitação, buscou uma solução cosmológica para tal. Para que o seu universo fizesse sentido naquele tempo, adicionou a constante cosmológica nas suas equações para tê-lo estático em larga escala. Depois dos anos 20 do século passado, quando Edwin Hubble observou a expansão cósmica, a constante cosmológica tornou-se desnecessária (o universo não era mais estático), o que teria feito Einstein afirmar o seu maior erro. Entretanto, em 1998 com a observação de que o universo expande-se de forma acelerada, a constante cosmológica reapareceu na cosmologia. No modelo padrão, ela é identificada com a energia escura.

**290** | *cadernos Nietzsche* 32, 2013

Mas, talvez, o maior problema do modelo padrão cosmológico seja a singularidade inicial, também conhecida como *big bang*. Como apontam Novello e Bergliaffa, "(...) uma singularidade pode ser naturalmente considerada como uma fonte de ausência de leis (…), pois a descrição do espaço-tempo não funciona lá [na singularidade inicial], e as leis da física pressupõem espaço-tempo."[20] E singularidades surgem naturalmente na teoria da gravitação de Einstein, de acordo com os teoremas de singularidade de Stephen Hawking e Roger Penrose. Seja no problema cosmológico ou no interior de um buraco negro, uma singularidade se apresenta[21]. Quando voltamos no tempo, no chamado tempo inicial, de acordo com o modelo padrão da cosmologia e a sua geometria do tipo Friedmann-Lemaître-Robertson-Walker, temos a singularidade inicial, onde grandezas físicas como curvatura do espaço-tempo, temperatura e densidade de energia tendem a valores infinitos. Para eliminar esse problema — a singularidade inicial ou mais uma sombra do Deus morto[22] —, surgiram as chamadas cosmologias com ricochete (*bouncing cosmologies*)[23]. Dentro destas cosmologias não singulares, sem a singularidade inicial e fora do modelo padrão cosmológico[24], temos

---

20  Cf. NOVELLO, M. & BERGLIAFFA, S. E. P. Bouncing cosmologies, in *Physics reports*, vol. 463, 2008, p. 129.
21  Singularidades são limitações da teoria de Einstein. Para maior parte dos físicos, o entendimento e a eliminação dessas singularidades serão possíveis somente com a ainda não criada teoria quântica da gravitação.
22  No *A gaia ciência*, Nietzsche escreve: "Deus está morto; mas, tal como são os homens, durante séculos ainda haverá cavernas em que sua sombra será mostrada" (FW/GC 108, KSA 3.467. Trad. PCS).
23  As cosmologias com ricochete surgem em vários contextos: desde o contexto einsteiniano até os chamados mundos branas, onde o nosso universo com quatro dimensões (três dimensões espaciais mais o tempo) está imerso num espaço-tempo com cinco dimensões.
24  Por serem fora do padrão, as cosmologias com ricochete têm menos adeptos (mesmo depois da observação, em 1998, da expansão acelerada do universo, considerada efeito da energia escura, um tipo de energia exótica que possibilita a existência de ricochetes em cosmologias einsteinianas). O motivo apresentado por Novello e Bergliaffa para o pouco interesse dos físicos em relação às cosmologias com ricochete é a existência dos chamados teoremas de singularidade, de Hawking e Penrose. Além do motivo apresentado por Novello e Bergliaffa, acreditamos que no ocidente judaico-cristão a ideia de um tempo linear, um início e um fim as cosmologias cíclicas[25]. Nestas últimas, não há um início ou um fim para o tempo, não existe a necessidade de especificar as condições iniciais para o universo. Temos, então, dentro das cosmologias com ricochete, particularmente as cíclicas, modelos cosmológicos que apontam na direção de um eterno retorno. O tempo é infinito, e a quantidade de energia é finita (ou forças finitas na linguagem nietzschiana). O universo expande-se e contrai-se sucessivamente, eternamente; sofre ricochetes depois de cada contração, sem passar por um *big bang*, voltando a expandir-se. Nessa interpretação física, o universo cria-se e destrói-se eternamente.

Mas mesmo resolvendo o problema da singularidade inicial, removendo-a, excluindo mais uma sombra do Deus morto, as cosmologias cíclicas apresentam problemas. Um dos seus problemas é assegurar que "(...) a estrutura em larga escala presente em um ciclo (…) não seja extinta por perturbações ou estruturas geradas em ciclos anteriores e não interfira em estruturas geradas em ciclos posteriores."[26] Ou seja, estruturas como galáxias e aglomerados de galáxias, no ciclo atual, não devem ter sofrido interferências de ciclos anteriores e não devem interferir em ciclos posteriores a este que vivemos.

Ainda uma questão importante se coloca: nessas cosmologias cíclicas, o retorno é o do mesmo? Como já dissemos, apontam para um universo que se cria e se destrói e para uma eternidade imanente, mas ainda não nos dizem muito sobre o retorno do mesmo

---

do tempo, seja mais aceitável para a maioria. Cf. NOVELLO, M. & BERGLIAFFA, S. E. P. Bouncing cosmologies. In: *Physics reports*, vol. 463, 2008, p. 206.
25  Cosmologias oscilatórias ou cíclicas — onde o universo expande-se e contrai-se eternamente — foram inicialmente estudadas por Richard Tolman na década de trinta do século passado. Naquele tempo, o maior impedimento para essas cosmologias foi a segunda lei da termodinâmica. Acreditava-se que a entropia necessariamente aumentaria de um ciclo para outro. Hoje em dia, no entanto, existem modelos cosmológicos cíclicos que criam uma solução para esse problema, assim como para o aumento do tamanho de buracos negros de um ciclo para outro.
26  Cf. NOVELLO, M. & BERGLIAFFA, S. E. P. Bouncing cosmologies. In: *Physics reports*, vol. 463, 2008, p. 207.

porque o problema da influência de ciclos anteriores em ciclos posteriores está em aberto. Na nossa visão, o retorno do mesmo, o retorno de "tudo na mesma ordem e sequência", somente ocorrerá se ciclos iguais ocorrerem.

*4 – Por que "provar"?*

Uma interessante questão sobre o eterno retorno nietzschiano é a tentativa do seu autor em "prová-lo" cientificamente. Nietzsche teria planejado estudar, nos anos oitenta do século XIX, física e matemática numa universidade europeia para dar bases científicas ao retorno[27]. Por que "prová-lo"? Paolo D'Iorio afirma de forma jocosa que Nietzsche teria o intuito de fortalecer sua ideia e "retornar aos escritos filosóficos como o mestre do eterno retorno"[28]. Müller-Lauter pergunta: "Pretenderia o filósofo conduzir o espírito da época orientado pelas ciências naturais ao pensamento do retorno de modo inapropriado, a fim de promover o processo de aproximação gradativa da nova 'verdade'?"[29] E responde: "Contra isso, pode-se dizer, entre outras coisas, que suas anotações para as provas da doutrina encontram-se em escritos póstumos (...)"[30], negando-se, dessa forma, o caráter exotérico de tais "provas". Mas, talvez, por não possuir uma poderosa arma como o teorema de Poincaré ou uma cosmologia não singular, Nietzsche não tenha publicado em seus livros uma forte "prova" científica a favor do eterno retorno. Então, a resposta para a pergunta de Müller-Lauter poderia ser afirmativa. Uma aproximação do homem que valoriza a ciência à doutrina do eterno retorno seria facilitada com algum tipo de "prova" ou embasamento científico.

---

[27] Uma forma de corroborar o eterno retorno, para Nietzsche, era refutar a segunda lei da termodinâmica. Para o filósofo, essa lei, que pode conduzir à morte térmica do universo, é teleológica. Sua recusa a qualquer fim ou teleologia fica clara no fragmento seguinte: "Se o mundo tivesse um alvo, teria de estar alcançado (…). Se fosse em geral apto a perseverar, tornar-se rígido, apto a um 'ser', se em todo o seu vir-a-ser tivesse apenas por um único instante essa aptidão ao 'ser', mais uma vez, há muito teria terminado todo vir-a-ser (…)" (Nachlass/FP 1885, 36[15], KSA 11.556. Trad. RRTF).
[28] Cf. D'IORIO, P. The eternal return: genesis and interpretation. Trad. Frank Chouraqui. In: *Agonist: a Nietzsche circle journal*, vol. 4, 2011, p. 39.
[29] Cf. MÜLLER-LAUTER, W. *Nietzsche: sua filosofia dos antagonismos e os antagonismos de sua filosofia*. Trad. Clademir Araldi. São Paulo: Editora Unifesp, 2009, p. 264.
[30] Idem, Ibidem.

Entretanto, a ciência para Nietzsche não passa de uma ficção. Por ser uma ficção, não está fundada na verdade, mas está no erro, na ilusão. Já em *Humano, demasiado humano*, o filósofo critica a matemática e a sua base, a lógica ou o princípio da identidade: "A invenção das leis dos números se deu com base no erro, predominante já nos primórdios, segundo o qual existem coisas iguais (mas realmente não há nada igual), ou pelo menos existem coisas (mas não existe nenhuma 'coisa')" (MA I/HH I 19, KSA 2.40. Trad. PCS). No *A gaia ciência* prossegue e afirma que "a tendência predominante de tratar o que é semelhante como igual — uma tendência ilógica, pois nada é realmente igual — foi o que criou todo fundamento para a lógica" (FW/GC 111, KSA 3.71. Trad. PCS). Em *Além do bem e do mal*, escreve o que pensa sobre a física: "Começa a despontar em cinco, seis cérebros, talvez, a ideia de que também a física é uma interpretação e disposição do mundo (nisso nos acompanhando, permitam lembrar!), e não uma explicação do mundo (...)" (JGB/BM 14, KSA 5.28. Trad. PCS). Então, vemos que para Nietzsche uma "prova" científica do eterno retorno não seria considerada uma verdade no sentido que a metafísica atribui a essa palavra, isto é, como a perfeita adequação entre mundo e pensamento. Pensamos que uma "prova" científica deveria servir, para o filósofo, como isca para uma lenta aproximação do homem que valoriza a ciência à doutrina ou cosmologia do eterno retorno do mesmo. Além disso, na nossa interpretação, poderia também servir como um exemplo do perspectivismo nietzschiano, como mais um olhar para o problema — um olhar científico. Pois como Nietzsche escreve no *Genealogia da moral*, "(...) *quanto mais* afetos permitirmos falar sobre uma coisa, *quanto mais* olhos, diferentes olhos, soubermos utilizar para essa coisa, tanto mais completo será o nosso 'conceito' dela, nossa 'objetividade'" (GM/GM III, 12, KSA 5.365. Trad. PCS).

**Abstract**: In this work we have carried out an approach between the nonsingular scientific cosmologies (without the initial singularity, the big bang), specially the cyclic models, and the Nietzsche's thought of the eternal recurrence. Moreover, we have pointed out reasons for the Nietzsche's search for scientific proofs about the eternal recurrence in the decade of 1880's.
**Keywords**: science - cosmology – eternal recurrence – forces

## referências bibliográficas

BARRENECHEA, M.; FEITOSA, C.; PINHEIRO, P. & SUAREZ, R. (orgs). *Nietzsche e asciências*. Rio de Janeiro: Editora 7 Letras, 2011.
BRUSH, S. *The kind of motion we call heat: a history of kinetic theory of gases in the 19th century*. Amsterdam: Elsevier Science & Publishers B. V., 1976, vols. I e II.
D'IORIO, P. The eternal return: genesis and interpretation. Trad. Frank Chouraqui. In: *Agonist: a Nietzsche circle journal*, vol. 4, 2011.
ELIADE, M. *Mito do eterno retorno*. Trad. José A. Ceschin. São Paulo: Mercuryo, 1992.
JAMMER, M. *Concepts of force*. New York: Dover Publications Inc., 1999.
MARTON, S. *Nietzsche: das forças cósmicas aos valores humanos*. 3. ed. Belo Horizonte: Editora UFMG, 2010.
______. O eterno retorno do mesmo: tese cosmológica ou imperativo ético. In: *Extravagâncias: ensaios sobre a filosofia de Nietzsche*. 3. ed. São Paulo: Discurso Editorial/Editora Barcarolla, 2009.
______. Da biologia à física: vontade de potência e eterno retorno do mesmo. Nietzsche e as ciências da natureza. In: Miguel Barrenechea *et al*. (orgs). *Nietzsche e as ciências*. Rio de Janeiro: Editora 7 Letras, 2011.
MÜLLER-LAUTER, W. *Nietzsche: sua filosofia dos antagonismos e os antagonismos de sua filosofia*. Trad. Clademir Araldi. São Paulo: Editora Unifesp, 2009.

NIETZSCHE, F. *Humano, demasiado humano*. Trad. Paulo César de Souza. São Paulo: Companhia das Letras, 2005.

\_\_\_\_\_\_\_. *A gaia ciência*. Trad. Paulo César de Souza. São Paulo: Companhia das Letras, 2001.

\_\_\_\_\_\_\_. *Assim falou Zaratustra*. Trad. Paulo César de Souza. São Paulo: Companhia das Letras, 2011.

\_\_\_\_\_\_\_. *Além do bem e do mal*. Trad. Paulo César de Souza. São Paulo: Companhia das Letras, 1992.

\_\_\_\_\_\_\_. *Genealogia da moral*. Trad. Paulo César de Souza. São Paulo: Companhia das Letras, 1998.

\_\_\_\_\_\_\_. *Ecce Homo*. Trad. Paulo César de Souza. São Paulo: Companhia das Letras, 1995.

\_\_\_\_\_\_\_. *Fragmentos do espólio: primavera de 1884 a outono de 1855*. Trad. Flávio R. Kothe. Brasília: Editora Universidade de Brasília, 2008.

\_\_\_\_\_\_\_. *Fragmentos finais*. Trad. Flávio R. Kothe. Brasília: Editora Universidade de Brasília, 2002.

\_\_\_\_\_\_\_. *Obras incompletas*. Trad. Rubens Rodrigues Torres Filho. 3. ed. São Paulo: Abril Cultural, 1983.

NOVELLO, M. & BERGLIAFFA, S. E. P. Bouncing cosmologies. In: *Physics reports*, vol. 463, p. 127-213, 2008.

RUBIRA, L. *Nietzsche: do eterno retorno do mesmo à transvaloração de todos os valores*. São Paulo: Discurso Editorial/Editora Barcarolla, 2010.